# FROM THE EPICYCLES OF THE GREEKS TO KEPLER'S ELLIPSE THE BREAKDOWN OF THE CIRCLE PARADIGM

**Dino Boccaletti**
Dipartimento di Matematica "G. Castelnuovo"
Università degli Studi di Roma "La Sapienza"
Rome, Italy
boccaletti@uniroma1.it

## Abstract

The principle that celestial bodies must move on circular orbits or on paths resulting from the composition of circular orbits has been assumed as a constant guide in the astronomical thougth of the peoples facing the Mediterranean sea as from the second century B.C. until the beginning of the XVII century. The mathematical model based on such an assumption, the theory of epicycles in all its versions and modifications, has been taken as a scheme for all astronomical calculations during at least eighteen centuries, from Hipparcus to Kepler. Kepler, who in 1609 eventually put aside what, paraphrasing Kuhn, can be called the circle paradigm, at the beginning of his Astronomia Nova says:

"Planetarum motus orbiculares esse peremnitas testatur. Id ab experientia mutuata ratio statim praesumit gyros ipsorum perfectos esse circulos. nam ex figuris circulus, ex corporibus coelum, censentur perfectissima. Vbi vero diligenter attendentes experientia diversum docere videtur; quod Planetae a circuli simplici semita exorbitent; plurima existit admiratio, quae tandem in caussas inquirendas homines impulit."[1]

As it is known, in Astronomia Nova (1609), Kepler succeded to establish the two laws which after him were named the first and the second Kepler's laws. The revolution he performed by giving up the circle paradigm is of fundamental importance and represents the indispensable premise to Newtonian theory. This revolution, the result of what Kepler called his "war" against Mars, is usually underestimated if one considers the break carried out with respect to the pre-Kepler celestial kinematics. In "Astronomia Nova" we assist to the attempt of Kepler to get for Mars an orbit consistent with the results of Tycho Brahe's observations. Passing through several phases ("hypotesis vicaria", oval, …) and step by step discarding the hypoteses wich gave results in contrast with observations, Kepler arrived at establishing (Astronomia Nova, Chapter LIX) that the orbit of Mars around the Sun is an ellipse.

Here an esposition is given of this work of analysis by Kepler wich is, in the history of science, one of the early examples of rigorous setting up of a model which correctly explains the experimental results.

[1] Johannes Kepler, New Astronomy: "The testimony of the ages confirms that the motions of the planets are orbicular. It is an immediate presumption of reason, reflected in experience, that their gyrations are perfect circles. For among figures it is circles, and among the bodies the heavens, that are considered the most perfect. However when experience is seen to teach something deviate from a simple circular path, it gives rise to a powerful sense of wonder, which at length drives men to look into causes.", translated by William H. Donahue, Cambridge U. P., 1992, p. 115

## Introduction

My talk concerns Johannes Kepler and the significance and weight one must attach to his "revolution", which consisted in the introduction of elliptic orbits for the planetary motions. According to a noted scholar of him, "(Kepler) est, pourrait-on dire, à la fois en avance, et en retard sur ses contemporaines"[2]. In the sub-title of this talk I made use of the celebrated Kuhn's definition of paradigm. It is surprising that such a remarkable example of breaking off of a paradigm still extant after eighteen centuries was not considered by Kuhn with more attention. In fact he just observed that "Kepler's account of his prolonged struggle with the motion of Mars and Priestley's description of his response to the proliferation of new gases provide classic examples of the more random sort of research produced by the awareness of anomaly."[3]

Charles Sanders Peirce (1839-1914), the beginner of pragmatism in the American philosophy, was certainly of opposite opinion on the "randomness" of the introduction of elliptic orbits, since he went so far as to define it "the greatest piece of Inductive reasoning ever yet conducted"[4].

In order to set the "*long war*" of Kepler against Mars - "inobservabile sidus" according to Plinius's definition quoted by himself - in the right historical context, let us briefly outline the situation in the eighteen centuries preceding Kepler, starting from the Greeks of third century B.C.

## The theory of epicycles[5]

The Greeks formalized their ideas about the motion of celestial bodies in the *theory of eccentrics* and in the *theory of epicycles.* The fundamental fact to keep in mind is that according to the Greeks the celestial bodies could not move on any kind of curve unless it was a circle (the perfect curve and therefore the only one worthy of a celestial body). This implied that any motion could only be either a uniform circular motion or a combination of uniform circular motions. We should also add that, in so doing, they renounced the possibility of considering seriously the problem of investigating the true nature of the physical system of the world, namely of finding the causes of the motion of celestial bodies. The various mathematical elaborations were basically oriented towards the *description* of motions. Through a crude schematization, we can say that from Hipparchus (second century B.C.), through Ptolemy (second century A.D.), to Kepler (1609 is the year of publication of Astronomia Nova), and finally to Newton (the *Principia* was published in 1687) attention has been given mainly to the *kinematics* and not to the *dynamics of* celestial bodies.

The theory which almost certainly appeared first is that of the *eccentrics.* At the basis of this theory the Earth is situated at the centre of the universe and the Moon rotates around it on a circular orbit with a period of 27 days; the Sun also rotates around the Earth with a period of one year; the centre of these orbits does *not* coincide with the position of the Earth. The inner planets, Mercury and Venus, move on circles whose centres are always on the line joining the Sun and the Earth; the Earth remains outside these circles. The outer planets (Mars, Jupiter, Saturn) also move on circles whose centres are on the line joining the Earth and the Sun, but these circles have so great a radius as to always encircle both the Earth and the Sun. This theory was employed by Hipparchus to represent the motion of the Sun.

The *eccentric* (see Fig. 1) is a circle whose centre ($C$) does not coincide with the position of the observer[6] on the Earth ($E$). If the Sun ($S$) moves with uniform motion on this circle, and hence the angle $ACS$ increases uniformly, it will not be the same for the angle $AES$: this will increase more slowly when $S$ is close to $A$ (the apogee, or farthest point from the Earth) and more quickly when $S$ is close to $B$ (the perigee, or closest point to the Earth).

---

[2] A. Koyré: La Révolution Astronomique – Copernic Kepler Borelli, Hermann, Paris, 1961, p. 119
[3] Thomas S. Kuhn: The Structure of Scientific Revolutions, Second edition, University of Chicago Press, 1970, p. 82
[4] C. S. Peirce: The New Elements of Mathematics, edited by Carolyn Eisele, Vol. III/1, Mouton Publishers, The Hague-Paris, 1976, p. 170
[5] D. Boccaletti, G. Pucacco: Theory of Orbits, Vol. 1, Second Edition, Springer Verlag, 2001
[6] As the observer is in a geocentric system, motions should be reproduced as seen from the Earth.

The motion of the Sun could equally be obtained by means of the epicycle model, which dates back to Apollonius (end of the third century B.C.). In such a model, the body whose motion must be represented is considered to be moving uniformly on a circle (epicycle). The centre of the epicycle moves uniformly on a second circle called the *deferent*. To demonstrate the said equivalence, we shall refer to Fig. 2. If one takes as the deferent a circle equal to the eccentric but with its centre at $E$ (the dashed circle), and if one takes the point $S'$ on it such that $ES'$ is parallel to $CS$, then $S'S$ is equal and also parallel to $EC$. Then the Sun $S$, which moves uniformly on the eccentric, can be considered in the same way to be in uniform motion on a circle of radius $SS'$ whose centre $S'$ moves uniformly on the deferent. In this case, therefore, the two methods lead to the same result. The arrows in Fig. 2 show that the motion on the epicycle must occur in a direction opposite to that of the motion on the deferent.

Similar considerations apply for the motion of the Moon, although in this case the theory turns out to be very rudimentary and fails to give an account of the greater complexity of the lunar motion compared to the solar one. Hipparchus found a way to overcome these difficulties by taking an eccentric whose centre describes a circle around the Earth in a period of nearly 9 years (corresponding to the motion of the apses). The epicycle theory was later improved by using as a deferent an eccentric circle and by introducing a new point called the *equant*, placed symmetrically to the Earth with respect to the centre of the deferent. This point replaced the centre of the deferent as a point from which to see the centre of the epicycle moving with constant angular velocity. Somehow, this violated the rule that the considered circular motions should be uniform. Later still, the theory was improved by adding secondary epicycles, epicycles which had an epicycle as a deferent.

We shall not deal with this matter further, since our purpose has been only to capture the spirit of the epicycle theory. It would be the basis of the Ptolemaic system which would reach, substantially unchanged in its fundamental structures, the age of the Renaissance and beyond. It is well known, in fact, that, during the thirteen centuries following the composition of the *Almagest*, the models worked out by Ptolemy were modified and improved several times, particularly by the Arabian astronomers, and therefore instead of the Ptolemaic system one should speak of Ptolemaic systems; however, the basic concepts have never been a matter of controversy. The Earth was immobile at the centre of the world and all the other bodies moved around it with motions that, however complicated, would have to be explained as suitable compositions of circular motions.

## *The Copernican revolution and the circular motions*[7]

The idea of circular motion continued to obsess even Copernicus, who with his work *"De Revolutionibus orbium coelestium"* (published in Nuremberg in 1543, the year of his death) founded the new cosmology and established a new theory of planetary motion. The title of Chapter 4 of the first book of *De Revolutionibus* in fact reads: "Quod motus corporum coelestium sit aequalis ac circularis, perpetuus, uel ex circularibus compositus"[8] ("The reason why the motion of celestial bodies is uniform circular and perpetual or composed of circular motions"), and later on, with the need to employ rectilinear motion, he deemed it necessary to demonstrate that a rectilinear motion could be generated by the composition of two circular motions. By moving the centre of the system from the Earth to the Sun, Copernicus succeeded in obtaining for all the planets, as for the Earth, quite simple orbits (circular in a first approximation) and no longer curves like that of Figure 3, which represents the orbits (seen from the Earth!) obtained from the epicycle theory. Copernicus granted to the Earth, nevertheless, a prime role, by assuming in his planetary theories that the centre of the terrestrial orbit, rather than the barycentre of the Sun-planet system, was the centre of every motion. Of course, this led to mistakes which still required the use of epicycles as correctives in order to be in agreement with the results of the observations (albeit still rather inaccurate). This, however, did not worry the astronomers, who were used to a veritable crowd of epicycles; in fact, Copernicus, in that little work known as *Commentariolus*, which he had circulated handwritten and which seems to precede the editing of the *De Revolutionibus*, concluded almost in triumph: “And in this way, Mercury moves altogether by seven circles, Venus by five, the Earth by three and, around it, the Moon by four; at last Mars, Jupiter and Saturn, each by five. As a consequence, 34 circles are sufficient to explain the entire structure of the universe, as well as the dance of the planets”. Notwithstanding the error of having assumed the centre of the terrestrial orbit to be the centre of every motion (an error which, as later emphasized by Kepler, is considerable in the case of Mars’s motion), Copernicus marked the beginning of a new era for astronomy and for the determination of the orbits of celestial bodies. The world had to wait a little less than 70 years from the publication of *De Revolutionibus* to reach, with Kepler, the very high point of what we have called “the kinematics of celestial bodies” and at the same time the sunset of the epicycles.

[7] D. Boccaletti, G. Pucacco: Theory of Orbits, Vol. 1, Second Edition, Springer Verlag, 2001
[8] Nikolaus Kopernicus: Gesamtausgabe, Munich, 1949

However, the disappearance of the epicycles was only temporary: some time later, like the Phoenix, they rose again from their ashes. In fact, as remarked (perhaps for the first time) by G. V. Schiaparelli in the last century, Fourier series expansions brought the epicycles[9] back again, in modern dress, in celestial mechanics.

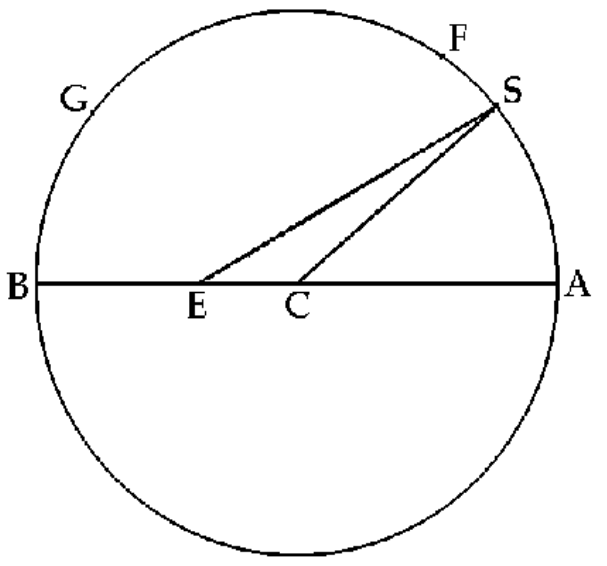

**Fig. 1**

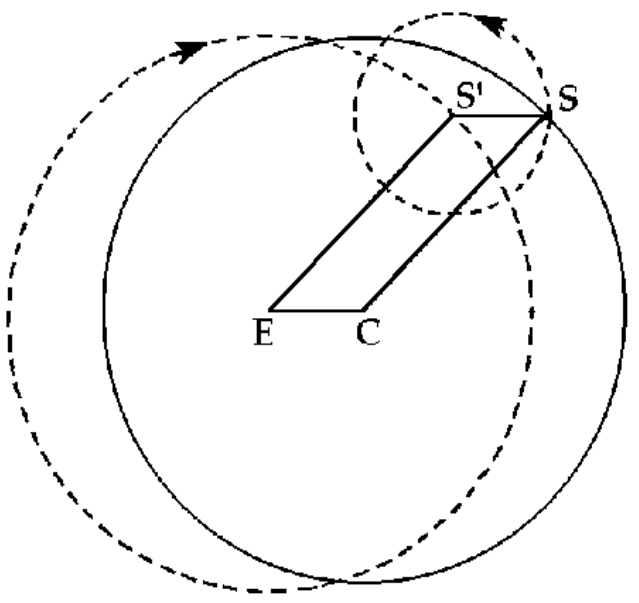

**Fig. 2**

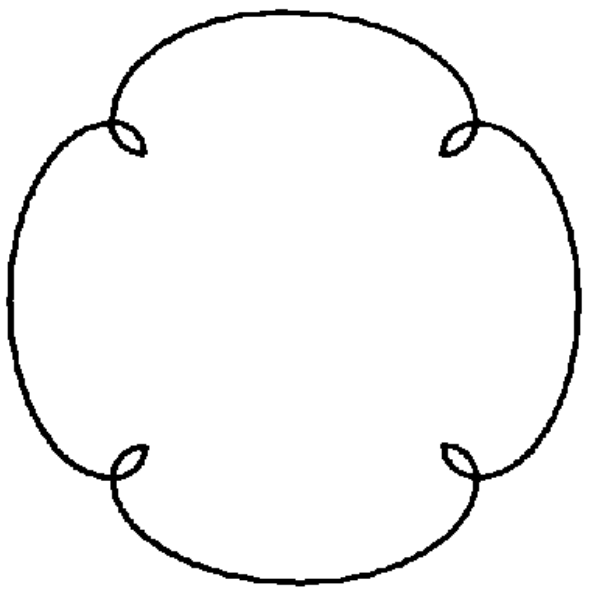

**Fig. 3**

## *Kepler establishes where the true centre of planetary motions is. The publication of Mysterium Cosmographicum*

After having received his master’s degree from the University of Tübingen on August 1591[10], Kepler entered the theological course. His intention was to become a priest, but halfway through his third year an event occurred that completely changed his life. Georgius Stadius, teacher of mathematics of the Lutheran School, died in Graz, and the local authorities asked the University of Tübingen for a new teacher. Kepler was chosen for this task, thus he interrupted the study of theology.

On 11 April 1594, Kepler arrived at Graz and took up his duties as teacher and provincial mathematician. He began to teach mathematics and astronomy. Since he had few pupils, he had much time to take the study of astronomy again. In 1596 this study resulted in the publication of the work which is usually called Mysterium Cosmographicum, though the complete title is extremely longer and each part of it has a well definite meaning (see Figure 4).

We give the reader an excerpt from the preface in which Kepler speaks about the genesis and the purpose of his work.

“Quo tempore Tubingae, ab hinc sexennio clarissimo viro M. Michaeli Maestlino operam dabam: motus multiplici incommoditate vsitatae de mundo opinionis, adeo delectatus sum

---

[9] See G. Schiaparelli: Scritti sulla Storia dell’Astronomia Antica, Parte Prima, Tomo Secondo, Zanichelli, Bologna, 1927, reprinted by Mimesis-IsIAO, Milano, 1997, p.335

[10] We cannot dwell upon Kepler’s life. We refer the reader to Max Caspar: Kepler, Dover Publications Inc., New York, 1993

Copernico, cuius ille in praelectionibus suis plurimam mentionem faciebat: vt non tantum crebro eius placita in physicis disputationibus candidatorum defenderem: sed etiam accuratam disputationem de motu primo, quod Terrae volutione accidat, conscriberem……Atque in hunc vsum partim ex ore Maestlini, partim meo Marte, quas Copernicus in Mathesi prae Ptolomaeo habet commoditates, paulatim collegi……Donec tandem Anno, etc. 95, cum ocium a lectionibus cuperem bene……toto animi impetu in hanc materiam incubui. Et tria potissimum erant, quorum ego causas, cur ita, non aliter essent, pertinaciter quaerebam, Numerus, Quantitas, et Motus Orbium."[11]

We cannot dwell upon his work on the whole, however it is enough to say that Kepler aimed chiefly at giving an explanation of the relations among the orbits of the planets known at that time. Kepler set up a complicated model in which planetary spheres are enclosed within the five platonic solids, starting with the cube between Saturn and Jupiter till to the octahedron between Venus and Mercury. But as far as our task is concerned, above all the clear exposition of the Copernican system is interesting. Quoting from Dreyer[12]:

"By two very instructive diagrams he shows that the Ptolemaic epicycles of the outer planets are seen exactly under the same angle from the earth as the orbit of the earth is from a point in each of the outer planetary orbits, and he shows how this explains why Mars has an epicycle of such enormous size, while that of Jupiter is much smaller and that of Saturn smaller still, though their excentrics are much larger than that of Mars. The Ptolemaic system could assign no cause for this curious arrangement, nor for the strange fact that the three planets when in

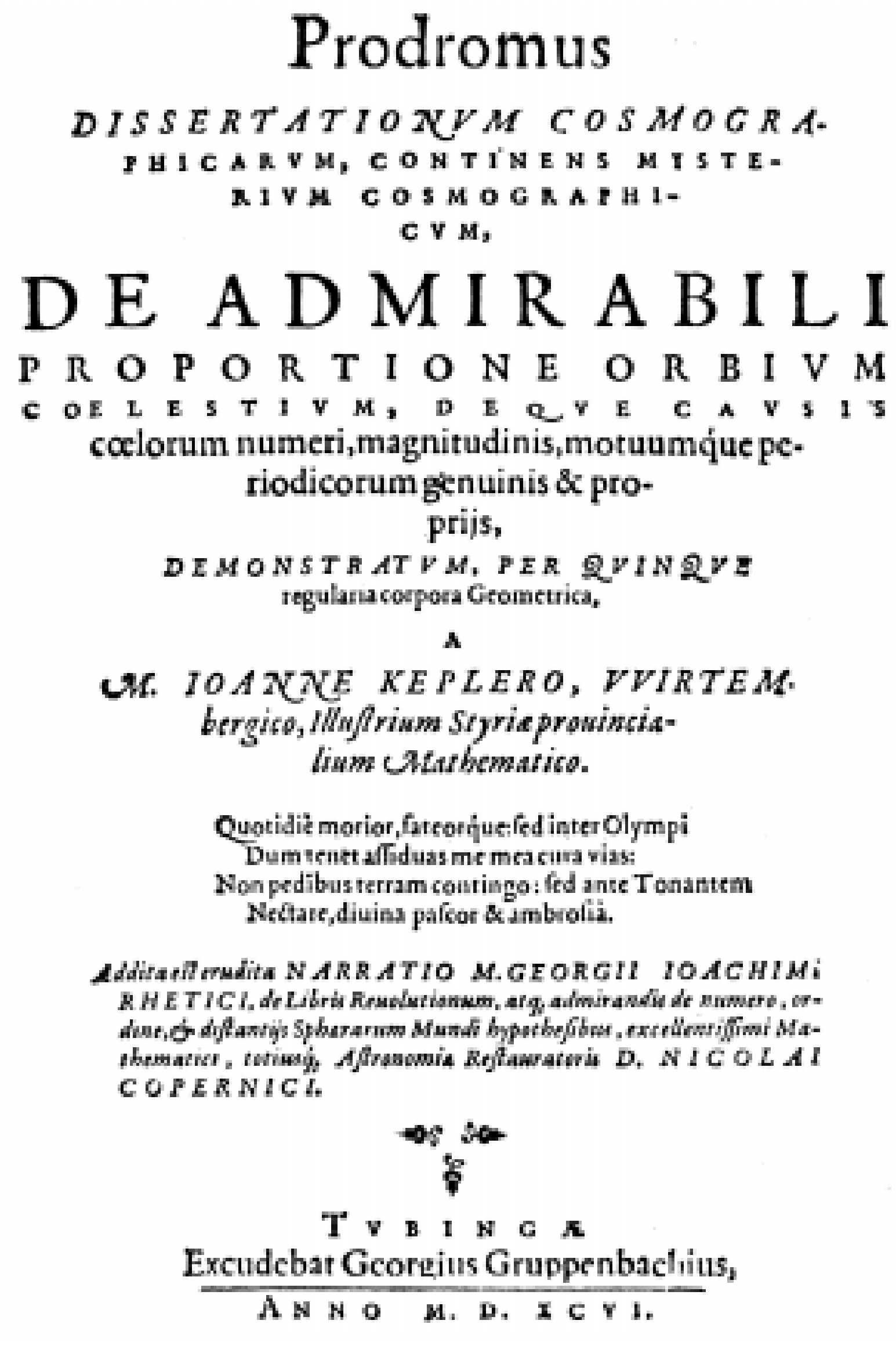

Prodromus

DISSERTATIONVM COSMOGRAPHICARVM, CONTINENS MYSTERIVM COSMOGRAPHICVM,

DE ADMIRABILI PROPORTIONE ORBIVM COELESTIVM, DE QVE CAVSIS cœlorum numeri, magnitudinis, motuumque periodicorum genuinis & proprijs,

DEMONSTRATVM, PER QVINQVE regularia corpora Geometrica,

A

M. IOANNE KEPLERO, VVIRTEMbergico, Illustrium Styriæ prouincialium Mathematico.

Quotidiè morior, fateorque: sed inter Olympi
Dum tenet assiduas me mea cura vias:
Non pedibus terram contingo: sed ante Tonantem
Nectare, diuina pascor & ambrosiâ.

Addita est erudita NARRATIO M. GEORGII IOACHIMI RHETICI, de Libris Reuolutionum, atq; admirandis de numero, ordine, & distantijs Sphærarum Mundi hypothesibus, excellentissimi Mathematici, totiusq; Astronomiæ Restauratoris D. NICOLAI COPERNICI.

TVBINGÆ
Excudebat Georgius Gruppenbachius,
ANNO M. D. XCVI.

**Fig. 4**

[11] "When I was studying under the distinguished Michael Maestlin at Tübingen six years ago, seeing the many inconveniences of the commonly accepted theory of the universe, I became so delighted with Copernicus, whom Maestlin often mentioned in his lectures, that I often defended his opinions in the students' debates about physics. I even wrote a painstaking disputation about the first motion, maintaining that it happens because of the rotation of the earth……I have by degrees – partly out of hearing Maestlin, partly by myself - collected all the advantages that Copernicus has over Ptolemy……At last in the year 1595 [in Graz] when I had an intermission in my lectures, I pondered on this subject with the whole energy of my mind. And there were three things above all for which I sought the causes as to why it was this way and not another - the number, the dimensions, and the motions of the orbs." The translation is from O. Gingerich: Kepler Johannes, in Dictionary of Scientific Biography, Charles Scribner's sons, New York, 1981

[12] J. L. E. Dreyer: History of Planetary Systems from Thales to Kepler, Cambridge University Press, 1906, p. 373

opposition to the sun should be in the perigees of their epicycles. Neither could it explain why the periods of the inner planets in their excentrics should be equal to that of the sun, nor give any reason why the sun and moon never became retrograde. All these facts are so simply explained by the doctrine of the earth's annual motion, while Copernicus is also able to account for precession without requiring "*that monstrous, huge and starless ninth sphere of the Alphonsines*." Certainly, it is difficult to see how anyone could read this chapter and still remain an adherent of the Ptolemaic system." (See Figures 5 and 6).

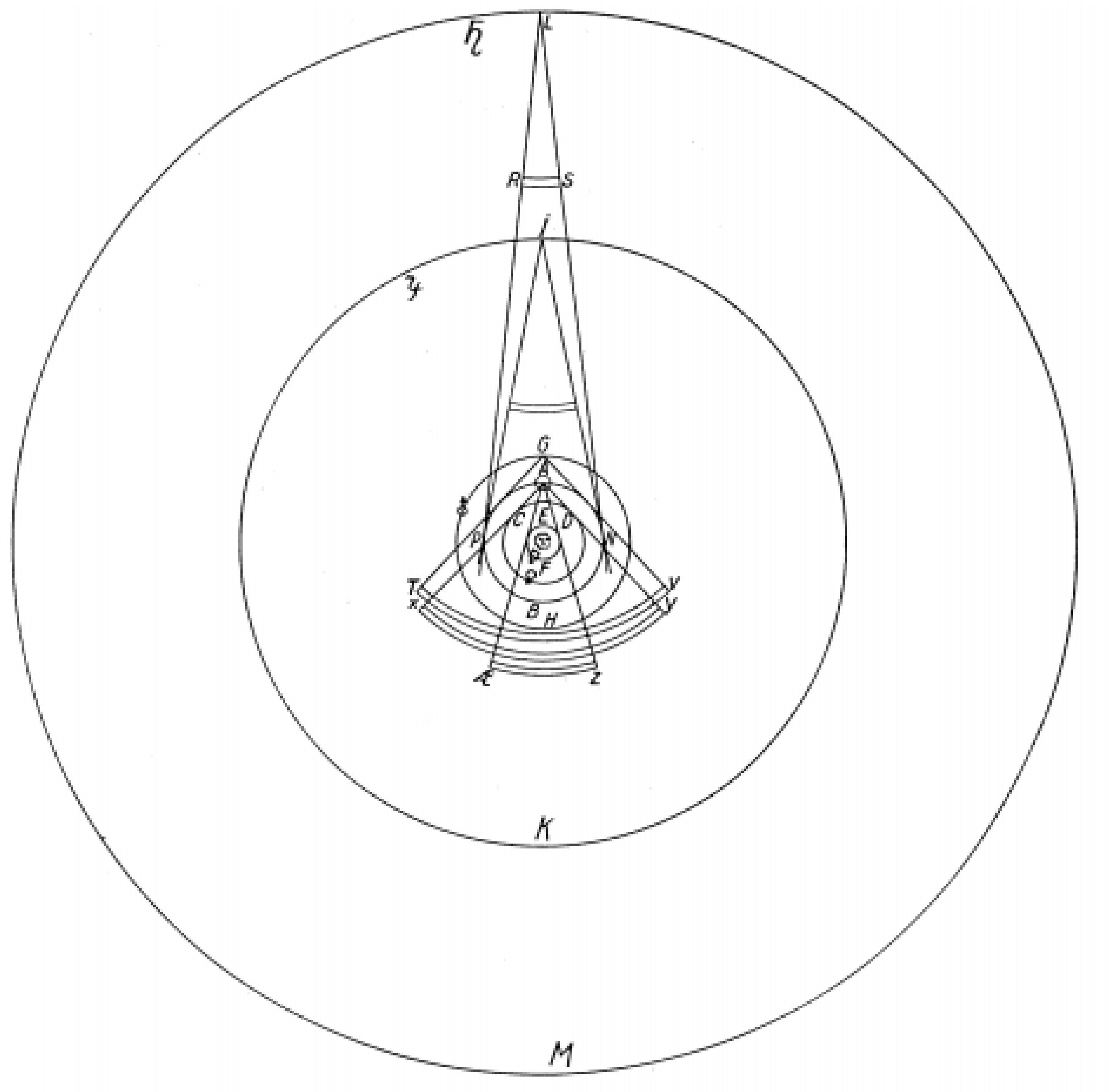

**Fig. 5**

Although Kepler's complicated geometrical model actually did not match with physical reality, anyhow it led Kepler to display a fundamental incongrousness of the Copernican theory. As we noted above, as a matter of fact Copernicus referred the orbits of the planets to the socalled "mean Sun" (that is, to the centre of the Earth's orb), rather than to the real Sun, even if he had placed the Sun at the centre of the universe. In fact, he wanted "to benefit by a shortening of calculations and not to trouble the painstaking lector of Ptolemy departing notably from him" [tamen ut calculum iuuet compendio, et ne nimium a Ptolemaeo recedendo, diligentem eius lectorem turbet][13]. Obviously, the two points (the mean Sun and the real Sun) are at a distance measured at any instant by the eccentricity.

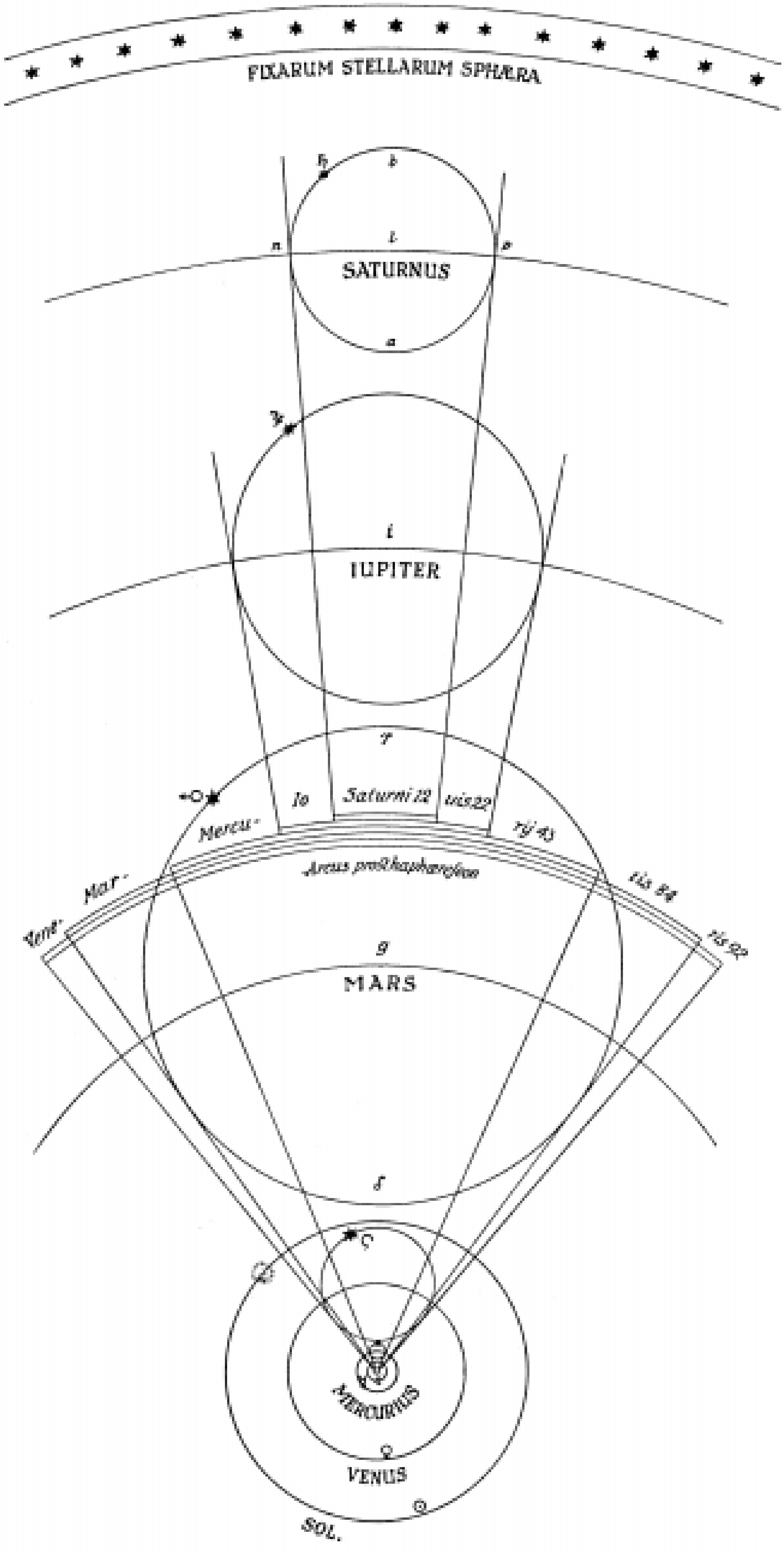

**Fig. 6**

Kepler wanted the distances of the planets from the centre of motions. The latter cannot be a merely geometric point, because the distances, which have to play a fundamental role in Kepler's model of universe, cannot be measured from an imaginary point. The origin of the motions of planets must be located uniquely in the body of the Sun, and not in the centre of the Earth's orb.

Paradoxically, the changes that Kepler made in Copernicus's model made it much more "Copernican". As a consequence, among other things, the Earth was truly considered in the same way of the other planets. In conclusion, Kepler removed from Copernicus's model of universe the last Ptolemaic remnants. Obviously, this required to recalculate all the distances (Kepler's old teacher Maestlin provided that), and introduced some discrepancies with previous results of observations. In addition, the shifting of the centre of motions from the mean Sun to the real Sun obliged later Kepler to introduce the equant (punctum aequans) again, which had been removed by Copernicus. Anyhow, Kepler did not changed his mind and waited for more reliable and precise observations. His encounter with Tycho Brahe resulted as fundamental with regard to this problem and influenced heavily his following work.

[13] Mysterium Cosmographicum, Chap. XV

We can conclude our few hints about Mysterium Cosmographicum with the words of Owen Gingerich: "Seldom in history has so wrong a book been so seminal in directing the future course of science"[14].

## *Astronomia Nova and the first two Kepler's laws*

Astronomia Nova, the second work of Kepler (published in Prague in 1609), is certainly the work that most intrigued both the contemporaries and the descendants. It is a work that does not resemble any other previous work, in both style and content. It has not the bombastic aspect of a scientific treatise; it looks rather like a diary in which the author notes day after day the results he obtained. But besides the results he obtained, there is also the passion with which the investigation is carried on and the little tricks adopted. All is mixed with religious-philosophycal considerations and autobiographical details. Moreover, certain chapters, previously written, are adapted to new results. Let us call upon one of the greatest historian of ancient astronomy to speak: "I wish to give…some examples of those rather trivial obstacles which every careful reader of Kepler's publications had to meet on practically every page. If Kepler ever did make any attempt to give a final polish to his writings, one can only say that he was not very successful. The number of trivial computing errors is enormous, parameters are changed without explanation (usually belonging to different stages of investigation, e. g. concerning the motion of the apsidal line of Mars), references to observations accessible to no one else are quoted sometimes in an incomplete form, sometimes for no evident reason, and so forth. To the historian of astronomy, this way of presentation is a blessing as great as a real disaster that leaves a city in total shambles is to the archaeologist"[15]. Of course, Neugebauer does not intend to diminish the greatness of Kepler, but only to forgive the readers contemporary with him: "……contemporaries cannot be blamed if their reaction was different from ours"[16]. Kepler inform us just in the title page (see Figure 7) that his work is "Plurium annorum pertinaci studium elaborata" (worked out in a tenacious study lasting many years[17]). Another fundamental element appearing in the subtitle is that Astronomia Nova is "based upon causes", i. e. it is Celestial Physics.

The aim Kepler proposed to himself is therefore not only to give a more rigorous kinematics to the motion of planets, but also to individuate the cause of the motion itself. The latter objective was not hit, and it was necessary to wait for Newton's work. Anyhow, the result that the motion of planets is due to an action exerted by the Sun comes from Kepler, even if Kepler did not think of gravity but of a magnetic force. We do not deal with that, but only with the work (the "*war*" against Mars, as he himself called it) done to come to the formulation – still rough and approximative - of the First and Second Kepler's Laws, as they are called today.

It is common knowledge that beginning from the century after the publication of Astronomia Nova the demonstrations given by Kepler (including the comparison between hypotheses and observations) were not considered as unquestionable. Newton himself, for instance, considers that Kepler had "guessed" the elliptical form of the orbit: "……Kepler knew ye Orb to be not circular but oval & guest it to be elliptical……"[18]. After Newton, generations of scholars attempted to interpretate Kepler's fascinating work and its refined Latin[19]. Delambre was one of the most important scholars of Kepler's work. He describes Astronomia Nova as "……l'ouvrage le plus beau, le plus important de Képler……cette composition dont Lalande et Bailly ont donné des

---

[14] Owen Gingerich, loc. cit.
[15] O. Neugebauer: Notes on Kepler, Communications on Pure and Applied Mathemathics, Vol. XIV, 1961, pp. 593-597
[16] Ibidem
[17] Donahue's translation (see Note 1)
[18] H. W. Turnbull (ed.): The correspondence of Isaac Newton, II, Cambridge, 1960, pp. 436-437
[19] Fortunately for the reader, we have now the excellent translation of Donahue (see Note 1)

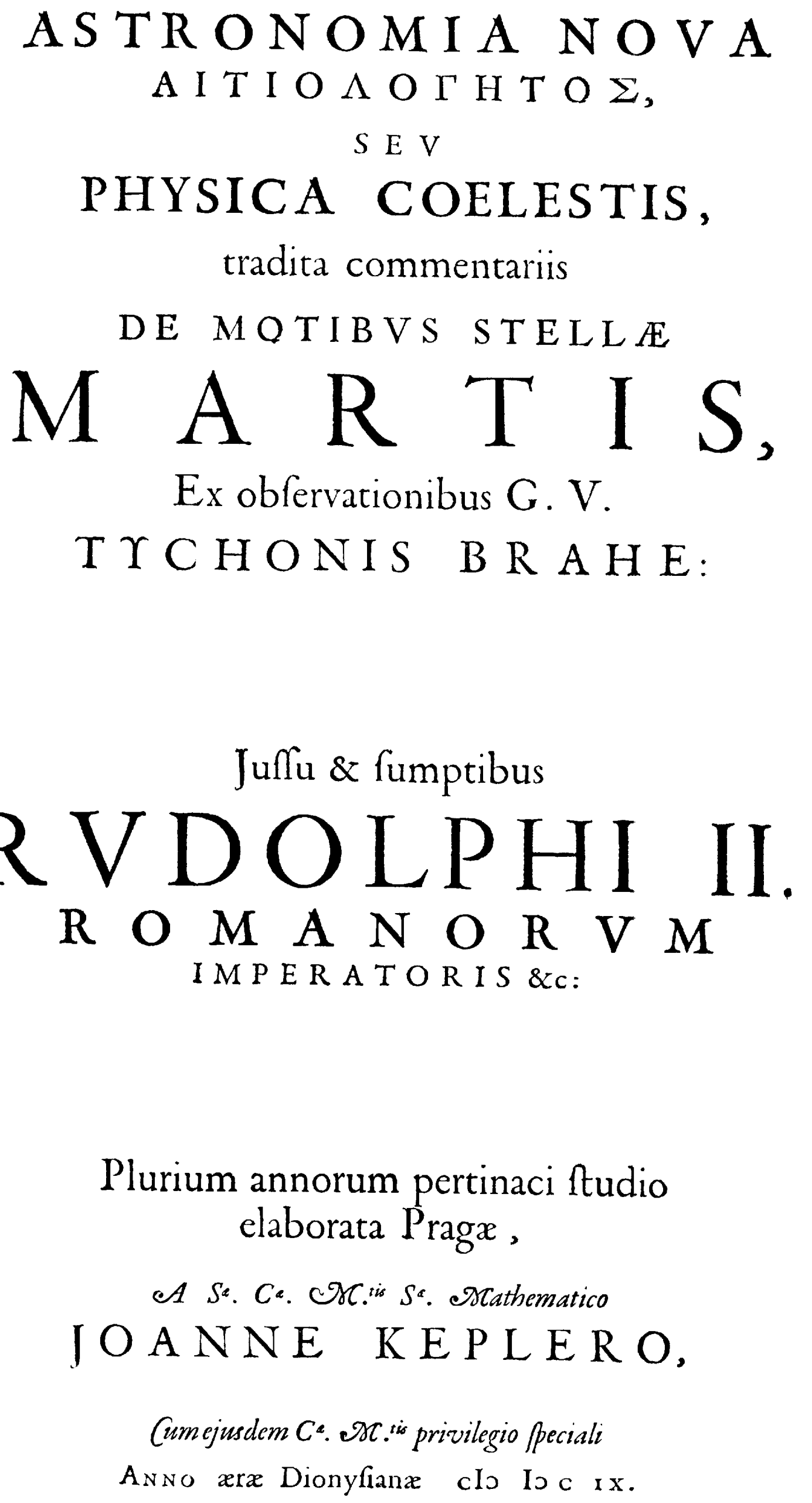

ASTRONOMIA NOVA
ΑΙΤΙΟΛΟΓΗΤΟΣ,
SEV
PHYSICA COELESTIS,
tradita commentariis
DE MOTIBVS STELLÆ
MARTIS,
Ex obſervationibus G. V.
TYCHONIS BRAHE:

Juſſu & ſumptibus
RVDOLPHI II.
ROMANORVM
IMPERATORIS &c:

Plurium annorum pertinaci ſtudio
elaborata Pragæ,

A S^e^. C^e^. M.^tis^ S^e^. Mathematico
JOANNE KEPLERO,

Cum ejusdem C^e^. M.^tis^ privilegio ſpeciali
ANNO æræ Dionyſianæ cIↃ IↃ c IX.

**Fig. 7**

extraits fort amples, mais qui sont loin d'être complets; et dont Lalande a dit que tout astronome la devait lire au moins une fois dans son entier"[20]. Notwithstanding this, Delambre did not spare Kepler his criticism with regard to the lack of evidence and the presence of mistakes, as we will see later. Another great Kepler's scholar, Robert Small[21], is nearly contemporary with Delambre. Almost a century later, we can remember the above cited Dreyer and also A. Berry[22], then passing to our contemporaries Owen Gingerich, Curtis Wilson, E. G. Aiton, Bruce Stephenson[23].

Neverthless, unlike the above cited authoritative scholars, we do not intend to enter into a technical discussion. Even if in our exposition we follow the line drawn by Curtis Wilson in his works, our intention is not of attempting to introduce new technical or historical-philosophical elements. Indeed we want to emphasize the revolutionary innovation introduced by the elliptical orbit in a scientific world where during eighteen century only circular orbits had been handled. The character of rupture is even more stressed by the fact that this revolution has not been received in its real significance for long time and that it was accepted only later as an empirical proposition (see the above cited Newton's opinion). As everybody knows, Kepler began to work at the theory of Mars's motion in 1600. During his studies he used the results of Tycho Brahe's observations, far the most precise of those available at that time. Let us mention that Tycho had at his disposal, when he was still living in Denmark, a personal observatory equipped with the most precise instruments ever constructed before the telescope. He had left Denmark in 1597 and two years later he had settled in Bohemia. Kepler paid him a first visit in January 1600. The following year he became his collaborator and then succeeded him to the office of imperial mathematician in 1601. Thus, Astronomia Nova and the “two laws” were elaborated during the first part (1600-1612) of Kepler’s staying in Prague. The work was published after complicated vicissitudes, including the controversies with Tycho Brahe’s heirs in 1609. When Kepler joined Tycho the first time in 1600, an opposition of Mars to the Sun had just taken place and a table of oppositions had been elaborated beginning from 1580. Tycho had also a very good theory at his disposal, by which he could represent the longitudes in opposition with an error of only 2’, but neither the latitudes nor the annual parallaxes. Kepler’s “war” against Mars started at that point. However, it is necessary to have arms to go to the war; Kepler had two arms, already set up in the Mysterium. The first one was the convincement that all planetary motions should be referred to the real Sun and not to the mean Sun; the second one was that all the planets, including the Earth, should move in the same way (around the real Sun). Maintaining the former argument, Kepler examines existing data on the oppositions of superior planets, and, according to Delambre, “……se jette dans un labyrinthe de calculs, qui ne sont pas de la dernière exactitude; il travaille sur des observations qui ne sont pas d’une grande précision; ses raisonnemens sont obscurs et ses conséquences incertaines……”[24]. Neverthless, Kepler obtained his first success, which consisted of removing the oscillations in the latitude bequeathed by the previous theories. Kepler’s thesis was that the plane of Mars’s orbit should have a constant inclination with respect to the plane of the ecliptic and should pass through the real Sun.

[20] Jean-Baptiste Joseph Delambre: Histoire de l'Astronomie Moderne, Tome Premier, Mme. Ve. Courcier, Paris, 1821, p. 389

[21] R. Small: An Account of the Astronomical Discoveries of Kepler, London, 1804 – Repr. Madison, Winsconsin, 1963

[22] A. Berry: A Short History of Astronomy – From earliest times through 19$^{th}$ century, 1898, Reprinted Dover 1961

[23] The list of Kepler's scholars is obviously very long and there are many important contributions. We just cite those who particularly devoted themselves to the subject of this paper. The reader can find exhaustive bibliographies in O. Gingerich (see Note 11), Curtis Wilson: Kepler's derivation of the elliptical path, Isis, **59**, 1968, pp. 5-25; From Kepler's Laws, So-called, to Universal Gravitation: Empirical Factors, Archive for History of Exact Sciences, **6**, 1970, pp. 89-170; E. J. Aiton: Kepler's Second Law of Planetary Motion, Isis, **60**, 1969, pp. 75-90; Bruce Stephenson: Kepler's Physical Astronomy, Princeton U. P., 1994; J. Evans, The history and practice of ancient astronomy, Oxford U. P., 1998. Finally, a wide bibliography was included in the Dover edition of Max Caspar's work – see Note 10.

According to Kepler, one of the reasons of complications in Copernicus’s theory of latitudes was the substitution of the real Sun by the mean Sun. He used three different methods to demonstrate his thesis. The second one was founded on observations concerning the particular situation in which the planet is in quadrature with the Sun, and both the Earth and the Sun lay on the line of nodes. In this case the observed latitude is equal to the inclination. By means of the three methods he found a constant inclination of 1°50’. So it was that the (apparent) oscillations in Mars’s orbits were eliminated. At that point Kepler proclaimed Copernicus “divitiarum suarum ipse ignarus”[25] (ignorant of his own riches).

The successive and more important step was the determination of the position of the line of apsides, of the eccentricity and of the mean anomaly at any epoch.

If we assume that a planet moves on a circular orbit around the Sun (Copernicus), we need three points to determine this orbit (in practice three observations performed at an opposition). Also for Ptolemy, who made use of the equant and assumed the Sun and the equant at the same distance from the centre of the circle and on the opposite side to it (bisection of the eccentricity), three oppositions were sufficient. On the contrary Kepler, who did not assume the bisection of the eccentricity even if using the equant, needed four oppositions. Among the ten oppositions observed by Tycho and the other two observed by himself in 1602 and 1604 (we recall that they happen about every 780 days) he chose the ones of 1587, 1591, 1593 and 1595 and deduced the exact time of opposition referred to the true Sun from them. At the exact time of opposition the Earth lies on a straight line between Mars and the Sun. Therefore, Mars is seen from the Earth against the background of the stars just as it would be seen by an observer standing on the Sun. Kepler was the first who calculated the oppositions with respect to the real Sun and not to the mean Sun. Those measurements provided the heliocentric longitude of Mars, i. e. its position on the ecliptic as it is seen from the Sun, but did not provide the distance from the Sun. To get the latter, an assumption is needed. At first, Kepler assumed a model of Ptolemaic type: Mars moves of uniform motion on a circular orbit around a point (the equant), which is not the centre of the circle; the Sun is off centre also. We called this model of Ptolemaic Type because the Ptolemaic configuration can be obtained from it simply replacing the Sun by the Earth. The unknown quantities are the line formed by the centre of Mars's orbit, the position of the Sun and the equant and the ratios of the distances between these points and the radius of Mars's orbit. As we reminded above, Kepler did not assume the bisection of the eccentricity, i. e. the same distances Sun-centre and centre-equant, but he wanted to place the centre of the orbit in the right position, between the Sun and the equant, in order to yield the best fit between observations and theory. He devoted the whole Chapter XVI of Astronomia Nova (fifteen "in folio" pages) to the exposition of the calculations performed to achieve the best fit. However, already at the fifth page he warned the reader: "Si te hujus laboriosae Methodi pertaesum fuerit, jure mei te misereat, qui eam ad minimum septuagies ivi cum plurima temporis jactura, & mirari defines hunc quintum jam annum abire, ex quo Martem aggressus sum, quamvis annus MCIII pene totus opticis inquisitionibus fuit traductus."[26] We show in Figure 8 the reproduction of one of the diagrams of Chapter XVI, where *d*, *e*, *f* and *g* represent the four positions of Mars at the four oppositions, *a*, *b* and *c* represent the Sun, the centre, and the equant, respectively, and *ih* is the line of apsides.

[24] Delambre, op. cit., p. 397
[25] Astronomia Nova, XIV, p. 81

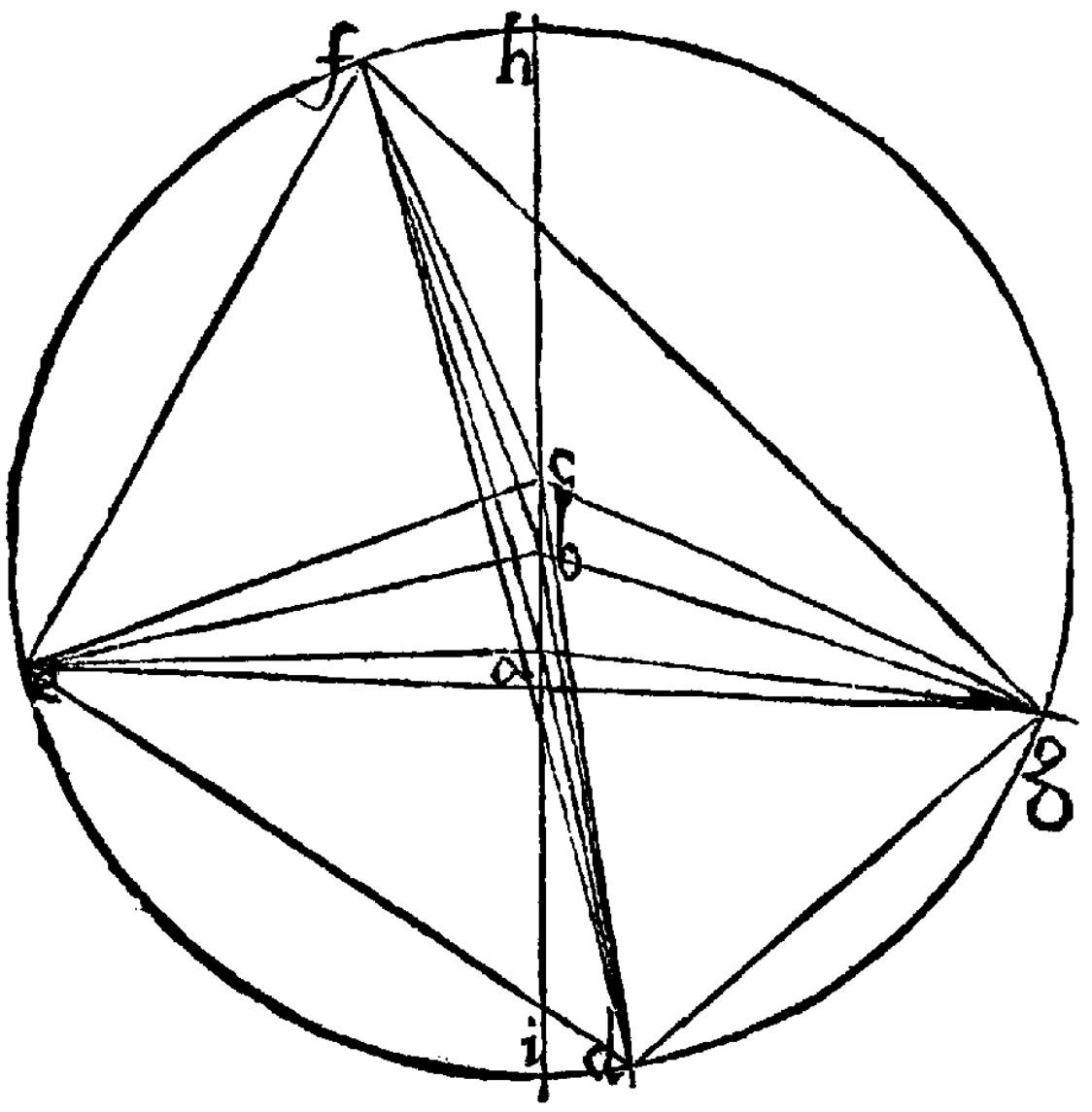

**Fig. 8**

The result of these wearisome calculations was for Kepler the heliocentric longitude of Mars with a precision comparable with that of observations, being 2'12'' the maximum discrepancy. Neverthless, the assumed model, later called "*hypothesis vicaria*", failed to give the correct distance between Mars and the Sun. This distance can be obtained by considering the triangle Sun-Earth-Mars, obviously when they are not in opposition. By means of the hypothesis vicaria, of observations and of Tycho's theory of solar motion (turned into a theory of Earth's motion through a simple geometrical transformation), one can calculate the positions of the sides Sun-Mars, Mars-Earth and Earth-Sun, respectively. The angles of this triangle can be obtained with a precision of four minutes of arc. Therefore, the ratios of the sides can be calculated trigonometrically. If one repeats this calculation for two observations, i. e. for two triangles, and makes use of Tycho's theory for the ratios of the two distances Earth-Sun (practically equal), the ratios of the corresponding distances Mars-Sun can be determined. By calculating the distance Mars-Sun when Mars was near the line of apsides, Kepler found out that the centre of Mars's orbit must be located nearly halfway between the Sun and the equant, whereas the best fit obtained through the hypothesis vicaria had given about 0.61 times the distance. On the other hand, if one took as good the value of about 0.5 (bisection of the eccentricity à la Ptolemy), one obtained the eccentrical longitude with an error of 8'. But Kepler could not consent to so great an error: "Nobis cum divina benignitas Tychonem Brahe observatorem diligentissimum concesserit, cujus ex observatis error hujus calculi Ptolemaici VII minutorum in Marte arguitur; aequum est, ut grata mente hoc Dei beneficium & agnoscamus & excolamus……Nunc quia contemni non potuerunt, sola igitur haec octo minuta viam praeiverunt ad totam Astronomiam reformandam, suntque materia magnae parti hujus operis facta."[27] Therefore, the trust in the reliability of Tycho's results and his

[26] "If the wearisome method has filled you with loathing, it should more properly fill you with compassion for me, as I have gone through it at least seventy times at the expense of a great deal of time, and you will cease to wonder that the fifth year has now gone by since I took up Mars, although the year 1603 was nearly all taken up by optical investigations." (Donahue's translation)

[27] Astronomia Nova, Chap. XIX, p. 113-114. "Since the divine benevolence has wouchsafed us Tycho Brahe, a most diligent observer, from whose observations the 8' error in Ptolemaic computation is shown, it is fitting that we with

obstinacy led Kepler to look for new ways. He had two theories at his disposal: Tycho's theory (turned into a theory of Earth's motion) and the hypothesis vicaria for Mars. Both made use of circular orbits with uniform angular motions around a point within the circle. Both gave, for the correspondent planet, the correct heliocentric longitude within two minutes of arc. However, whereas Tycho's theory gave the correct Earth-Sun distance, the hypothesis vicaria gave the wrong distance Mars-Sun. The assumptions of the hypothesis vicaria were two: circular trajectory and uniform motion around the equant. One or the other or both had to be wrong. If one held the assumption of circular trajectory, the equant was subjected to oscillations, and this made no sense. Kepler began the study of planetary motion again, looking for rules which were solid in the same way for all planets. He arrived to establish an approximate rule, which is the ancestress of the area law. Exploiting the assumption of the bisection of eccentricity, he arrived to maintain that the velocity of a planet varies inversely with respect to its distance from the Sun.

Although this is true only at perihelion and aphelion, Kepler generalized the rule, extending it to the whole orbit and to whatsoever planet. We used the term "velocity", but this is not exact because Kepler did not have the idea of instantaneous velocity as we intend it. He always spoke of the time ("*mora*" in Kepler's Latin) necessary to cover a given arc. For him "the elements of time and space were infinitesimal magnitudes but never zero, so that he envisaged time not as a flowing instant, but as a succession of durations" (Aiton). Reformulated in the light of these considerations, the rule becomes: the times necessary to cover small equal arcs are approximately proportional to the distances of these arcs from the Sun. This rule worked well for the Earth's orbit and agreed with Tycho's theory. Kepler took a further step toward what we now call the second law. He divided the semicircle (remember that he was still using circular orbits) into 180 arcs of one degree each, evaluated their distances from the Sun and, by summing the distances comprised in a given sector, put this sum proportional to the time elapsed to sweep the sector itself. After that, he replaced the sum of the distances comprised within a sector by the area of the sector itself, as if those segments were thick. In this way the area law was born. Its application to Earth's motion works well enough, with a discrepancy of only 34'' (remember that Earth's orbit has a very small eccentricity!). On the contrary, for Mars the discrepancy was remarkable. Here the most difficult part of Kepler's work began and also did the incubation of the revolutionary idea. Kepler still considered Mars's orbit as a circular one, with the Sun off centre. The Sun pushes the planet around by a "*species immateriata*", an immaterial virtue, which weakens in proportion to the distance. This point, still deserving to be deeply discussed, has been recently re-examined by Bruce Stephenson[28], who rejected the current interpretation. However, this goes beyond the scope of our talk. Kepler studied the complete orbit of Mars during the whole period (687 days) of its motion and compared the outcome of the area law with the predictions of the hypothesis vicaria. The results of the two theories coincided in the apsides and in the quadrants, but not in the octants (with a discrepancy of about 8'). By assuming the circularity of the orbit and the area law, one obtained that the planet moved too rapidly about the apsides and too slowly about the quadrants. Then either the circular orbit was wrong or the area law, or both. In the first case, one had to reduce the area near the quadrants, and to shorten the times for the given arcs accordingly, i. e. to make the orbit be oval. In Chapter XLV of Astronomia Nova the study begins "De causis naturalibus hujus deflectionis planetae a circulo", i. e. on the causes of the non-circularity of the orbit.

At that time, the great difficulties which Kepler met with concerned the quadrature of the ovoidal orbit. At the end, to simplify the computations, he replaced the oval by an ellipse he called "auxiliary ellipse" (it was not yet the true ellipse and the Sun was not in one of the two foci). After very cumbersome calculations concerning the "moonlet" comprised between Mars's orbit and the eccentric circle, he arrived to demonstrate on the basis of the area law that the true orbit deviates

---

thankful mind both acknowledge and honour this benefit of God……Now, because they could not have been ignored, these eight minutes alone will have led the way to the reformation of all of astronomy, and have constituted the material for a great part of the present work." (Donahue's translation) very little from the auxiliary ellipse. Finally, he realized that, replacing everywhere the radius vector of the eccentric circle by the same quantity times the cosine of the optical equation ("*distantia diametralis*", as Kepler called it), he obtained a complete agreement with the distances resulting from Tycho's observations.

[28] Bruce Stephenson, op. cit., pp. 68-69

He had discovered that the radius vector of Mars is always given by the equation

$$r = a + ae\cos E$$

$a$ being the radius of the eccentric circle, $e$ the eccentricity and $E$ the eccentric anomaly (measured from the aphelion after the ancient fashion). Obviously, it is easy at present time to assert that Kepler had already obtained the correct result and the orbit was elliptic with the Sun at one of the foci. However, Kepler still continued to worrying with circles before definitively surrending himself to the elliptic motion. He concluded Chapter LVIII finally admitting "O me ridiculum! perinde quasi libratio in diametro, non possit esse via ad ellipsin. Itaque non parvo mihi constitit ista notitia, juxta librationem consistere ellipsin; ut sequenti capite patescet: ubi simul etiam demonstrabitur, nullam Planetae relinqui figuram Orbitae, praeterquam perfecte ellipticam; conspirantibus rationibus, a principiis Physicis, derivatis, cum experientia observationum & hypoteseos vicariae hoc capite allegata."[29] And Chapter LIX is the chapter of the triumph: all the pictures are surmounted by the triumphal car (see Figure 9, where a picture from p. 286 of Astronomia Nova is reproduced). In this chapter, the first and the second law (as we call them today) are demonstrated and connected one to the other. But, as noted by Koyré, "aucun autre peut-être n'est plus embarassé, touffu - et confus – que celui-ci"[30].

In Chapter LIX Kepler explains the theorems on the properties of conics which he needs for the "physics" to be worked out in Chapter LX. At the end, with the consciousness of having dealed with a somewhat complicated matter, he apologizes to the reader by saying that he would have met with the same difficulties in reading Apollonius's treatise. But then, before this passage there is an assertion showing all Kepler's legitimate pride for the achieved success and his awareness that it was due to right physical principles and mathematical rigour. "Quod nisi causae physicae, initio a me susceptae, loco principiorum, probae essent nunquam in tanta subtilitate inquisitionis consistere potuissent."[31] As we have striven to expound so far, in Astronomia Nova one would look in vain for a formulation (as the one we use today) of the two laws. Kepler was more outspoken in the fifth book of Epitome, where the foci of the ellipse are called with this name (invented by himself in his work on optics), but it is completely evident from his conclusions that in Astronomia Nova the circle paradigm is definitively broken. We can add that it was not an easy undertaking.

Besides the hindrance due to the almost sacred nature of the circular model, there was the real difficulty of discriminating between the two models (elliptic orbit and area law versus eccentric circle with the equant). As J. Evans shows with great lucidity in the last pages of his book[32], it was a dilemma not easy to be solved.

---

[29] "O ridicolous me! To think that the reciprocation on the diameter could not be the way to the ellipse! So it came to me as no small revelation that through the reciprocation an ellipse was generated. This will be made clear in the following chapter, where it will be demonstrated at the same time, through the agreement of arguments from physical principles with the body of experience, mentioned in this chapter, that is contained in the observations and in the vicarious hypothesis, that no figure is left for the planet to follow other than a perfectly elliptical one." (Donahue's translation)
[30] A. Koyré, op. cit., p. 267
[31] "And unless the physical causes that I had taken in the place of principles had been good ones, they would never have been able to withstand an investigation of such exactitude." (Donahue's translation)
[32] J. Evans, op. cit., pp. 441-443

CAP. LIX.

# PROTHEOREMATA.

I.

SI intra circulum deſcribatur ellipſis, tangens verticibus circulum, in punctis oppoſitis; & per centrum & puncta contactuum ducatur diameter; deinde a punctis aliis circumferentiæ circuli ducantur per pendiculares in hanc diametrum: eæ omnes a circumferentia ellipſeos ſecabuntur in eandem proportionem.

*Ex l.* I. *Apollonii Conicorum pag.* XXI. *demonſtrat* COMMANDINVS *in commentario ſuper* V. *Sphæroideon* ARCHIMEDIS.

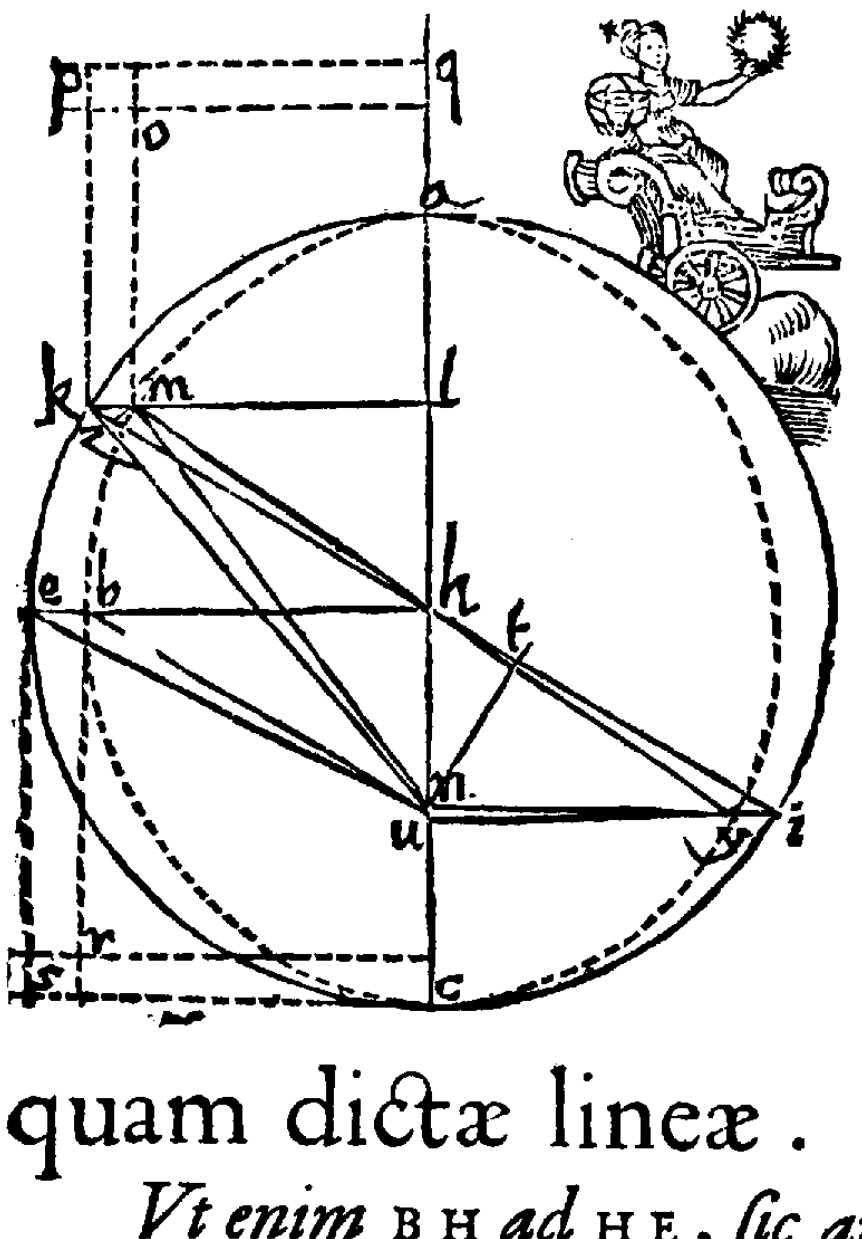

*Sit enim circulus* A E C. *in eo ellipſis* A B C *tangens circulum in* A C. *& ducatur diameter per* A. C. *puncta contactuum, & per* H *centrum. Deinde ex punctis circumferentiæ* K. E. *deſcendant perpendiculares* K L, E H, *ſectæ in* M. B. *a circumferentia ellipſeos. Erit ut* B H *ad* H E, *ſic* M L *ad* L K. *& ſic omnes aliæ perpendiculares.*

II.

Area ellipſis ſic inſcriptæ circulo, ad aream circuli, habet proportionem eandem, quam dictæ lineæ.

*Vt enim* B H *ad* H E, *ſic area ellipſeos* A B C *ad aream circuli* A E C. *Eſt quinta Sphæroideon* ARCHIMEDIS.

Fig. 9